\documentclass[structabstract]{aa}  

\usepackage{graphicx}
\usepackage{natbib}
\usepackage{url}
\usepackage{txfonts}
\newcommand{\jb}{{\sl J}}
\newcommand{\hb}{{\sl H}}
\newcommand{\kb}{{\sl K}$_S$}
\newcommand{\teff}{$T_{\rm eff}$}

\begin{document}

\title{Empirical Near Infrared colors for low-mass stars and brown dwarfs in the Orion Nebula Cluster}
\subtitle{An empirical Near Infrared isochrone at $\sim$1~Myr}

   \author{G. Scandariato\inst{1,4}
          \and
          N. Da Rio\inst{2}
          \and
          M. Robberto\inst{3}
          \and
          I. Pagano\inst{4}
          \and
          K. Stassun\inst{5,6}
          }

   \institute{Dipartimento di Fisica e Astronomia, Universit\`a di Catania, Italy\\
   \email{gas@oact.inaf.it}
         \and
             European Space Agency (ESTEC), PO\ Box 299, 2200 AG\ Noordwijk, The Netherlands\\
         \and
             Space Telescope Science Institute, Baltimore, MD 21218\\
         \and
          	INAF Osservatorio Astrofisico di Catania, via S.\ Sofia 78, 95123 Catania, Italy\\
		\and
			Department of Physics \& Astronomy, Vanderbilt University, Nashville, TN 37235, USA\\
		\and
			Department of Physics, Fisk University, Nashville, TN 37208, USA\\
             }


 \abstract
 {
Current atmospheric and evolutionary models for low-mass stars and brown dwarfs rely on approximate assumptions on the physics of the stellar structure and the atmospheric radiative transfer. This leads to biased theoretical predictions on the photospheric Spectral Energy Distributions of these system, especially when applied to low surface gravity objects such as Pre-Main Sequence (PMS) stars, and affects the derivation of stellar parameters from photometric data. 
}
 {
Our main goal is to correct the biases present in the theoretical predictions for the near-IR photometry of low-mass PMS stars. Using empirical intrinsic IR colors, we assess the accuracy of current synthetic spectral libraries and evolutionary models. We investigate how the uncertainty in the intrinsic colors associated with different PMS models affect the derivation of the Initial Mass Function of young clusters from near-IR photometry. 
}
 {
We consider a sample of $\sim$300 PMS stars in the Orion Nebula Cluster (age$\simeq$1~Myr) with well measured luminosities, temperatures and photospheric \jb\hb\kb\ photometry. This sample is used as a benchmark for testing both atmospheric and evolutionary theoretical models.  
}
 {
By analyzing the photospheric colors of our sample of young stars, we find that the synthetic \jb\hb\kb\ photometry provided by theoretical spectral templates for late spectral types ($>$K6) are accurate at the level of $\sim$0.2~mag, while colors are accurate at $\lesssim$0.1~mag. We tabulate the intrinsic photospheric colors, appropriate for the Orion Nebula Cluster, in the range K6-M8.5. They can be conveniently used as templates for the intrinsic colors of other young (age$\lesssim$5~Myr) stellar clusters.
}
 {
The theoretically-predicted \jb\hb\kb\ magnitudes of young late type stars do not accurately reproduce the intrinsic ones of the Orion Nebula Cluster members. An empirical correction of the atmospheric templates can fix the discrepancies between expected and observed colors. Still, other biases in the evolutionary models prevent a more robust comparison between observations and theoretical absolute magnitudes. In particular, PMS evolutionary models seem to consistently underestimate the intrinsic near-infrared flux at the very late spectral types, and this may introduce spurious features in the low-mass end of the photometrically-determined Initial Mass Function of young clusters.
}


   \maketitle

\section{Introduction}\label{sec:intro}

In the last two decades several generations of evolutionary models of pre-main-sequence (PMS) objects have been published with continuous improvements on the treatment of the hydrostatic structure, radiation/heat transfer and thermodynamic equilibrium. There are at least 7 different families of published PMS evolutionary calculations that have been widely circulated in machine-readable formats and that span a suitable range of stellar masses: \citet{Swenson1994}; \citet{Dantona97} with 1998 electronic-only update (DM98); \citet{palla99}; \citet{Siess00} (SDF00); \citet{BCAH98} (BCAH98) with the subsequent extensions to the substellar regime with the COND and DUSTY models \citep{Chabrier2000}; \citet{Yi2003}; and \citet{Tognelli11}. However, considerable differences still exist as the various sets of evolutionary tracks show systematic differences in the predicted masses and ages of stars in the HR diagram \citep{Hillenbrand2004,Hillenbrand2008}.

The comparison of theoretical predictions with observations on the HR diagram requires converting observed quantities, e.g.\ spectral type and some photometry in at least two passbands, into physical parameters like effective temperature and absolute luminosity \citep[see e.g.][]{Hillenbrand97}. This conversion is normally done using an empirical, or semi-empirical, calibration. It is also possible to take the complementary approach of deriving observational data from theoretical predictions. For example, plotting the photometry of a young stellar cluster in a color-magnitude (CMD) or two-color diagram (2CD) may show structures (like e.g.\ the cluster isochrone) that can be compared with the results of evolutionary and synthetic photometry calculations. Varying the model parameters one can directly visualize the effects of metallicity, effective temperature scale, surface gravity, reddening law, accretion \citep[see e.g.][D10 hereafter]{Dario10}. 

No approach is without drawbacks. On the one hand, the empirical corrections needed to derive the physical parameters of PMS stars, like e.g.\ colors  and  bolometric corrections, may have been derived on samples of stars which may not be fully representative of the sample under examination. There is a standing tradition, for example, of using for PMS stars the intrinsic colors of Main Sequence stars \citep[e.g.][]{Kenyon95}. The long-known difference between the spectral type vs.\ temperature relations for dwarfs and giants represents another source of uncertainty for PMS stars, that many authors overcome (or mitigate) this uncertainty using the \lq\lq intermediate\rq\rq\ scale of \citet{Luhman99}, which has been found adequate for PMS stars in many studies \citep[e.g., ][]{Dario10}. On the other hand, the systematic uncertainties of evolutionary models may combine with those of the atmospheric templates and provide erroneous results. The selection of adequate model atmospheres for PMS stars is especially critical for late type stars (M-type and later), which represent the peak of the Initial Mass Function \citep{Bastian2010} and whose spectral energy distributions (SEDs) are dominated by broad molecular absorption features. Cool atmospheres (\teff$\lesssim$3000~K) host a variety of molecules and provide an environment for dust condensation, whose role in the heat and radiation transfer through the photosphere cannot be neglected \citep{AHATS01, Allard10}. Moreover, for sub-stellar objects, convection may involve the photosphere, which therefore can no longer be treated as a system in radiative equilibrium. It is therefore crucial, particularly for cool atmospheres, to include in the synthesis codes the largest number of molecular lines in order to accurately reproduce the radiation transfer through the atmosphere. 

These difficulties are aggravated by observational limitations. It is hard to obtain high quality data for a statistically significant sample of PMS stars of comparable age and distance. Even in the solar vicinity, rich and young stellar clusters are typically affected by large and inhomogeneous reddening, being still enshrouded in their parental molecular cloud \citep{Lada03}, while the stars often show accretion excess. Older systems tend to be spatially spread and affected by membership uncertainties.  In this respect, due to its relatively low foreground extinction \citep[$A_V\lesssim 3$,][]{Scandariato10} and  vicinity \citep[$d=414$~pc,][]{Menten07}, the Orion Nebula Cluster (ONC) provides a unique opportunity to analyze the intrinsic colors of its members and assess the accuracy of PMS models. 

In the optical wavelength range, a recent attempt to calibrate empirically the colors of PMS stars is presented in \citet{Dario09}. In that work, based on optical BVI photometry of the ONC, the authors show how present families of synthetic spectra fail in matching the observed colors, and present a correction based on their data. In D10, they further refine this calibration, limited to the $I$ band and 2 medium bands at $\lambda\sim$7700~\AA, by explicitly calibrating colors as a function of \teff, and by decreasing the lowest \teff\ limit down to $\sim$2800~K.

The goal of this paper is to use the extensive set of spectral types and photometric data available for the ONC to test the theoretical models and to empirically determine the intrinsic (photospheric) \jb\hb\kb\ magnitudes and colors of the cluster members as functions of \teff. The intrinsic NIR colors of PMS stars in Orion we derive will be also appropriate for the (possibly ideal) cluster isochrone, and applicable to young (age$\lesssim$5~Myr) systems in general. In Sect.~\ref{sec:dataset} we present the selection of our sample of stars, based on the latest spectral characterization and our recently published NIR photometry. In Sect.~\ref{sec:analysis} we refine our list to select the subset of stars most suitable for our purposes. By means of this sample, in Sect.~\ref{sec:atm} we test the most recent atmospheric model of \citet{Allard10}, and in Sect.~\ref{sec:empiric} we derive the average colors of the cluster. Finally, in Sect.~\ref{sec:discussion} we discuss our result and we compare them to the current theoretical predictions.

\section{The data set}\label{sec:dataset}

We base our analysis on the ONC NIR photometric catalog of \citet{paperI}. This catalog contains \jb\hb\kb\ photometry in the 2MASS system for $\sim$6500 point-like sources, spread over an area of $\sim$30\arcmin$\times$40\arcmin\ roughly centered on $\theta^1$Ori-C (RA=$05^h 35^m16.46^s$, DEC=$-05^{\circ} 23\arcmin23.2\arcsec$), down to \jb\hb\kb$\sim$18. We cross-match the NIR catalog with the list of $BVI$ photometry of \citet{Dario09}. These NIR and optical catalogs are products of the HST Treasury Program on the Orion Nebula Cluster (HST GO-10246). They originate from \textit{simultaneous} observations of the cluster, and therefore are largely immune from the uncertainties introduced by source variability when multi-band photometry is collected at different epochs.

\citet[][D12 hereafter]{Dario11} provide photospheric luminosity, interstellar extinction and accretion luminosity for $\sim$1,200 stars, whose spectral types are known either from previous spectroscopic surveys \citep{Hillenbrand97} or from observations of the narrow-band photometric index of TiO, either new or from D10.

In order to extend our analysis down to later M-subtypes and probing the Brown Dwarfs (BDs) spectral range, we complement our list with the spectral classification of \citet[][R07 hereafter]{Riddick07} for 45 M-type stars not in our catalog. The authors provide extinctions, luminosities and spectral types, together with \jb\hb\kb\ photometry from their previous NIR photometric survey \citep{LR05}. 

We also add 51 more stars from the spectroscopic study of \citet[][S04 hereafter]{Slesnick04}, which also provides extinction, luminosity and spectral type of the stars. \hb\kb\ photometry is taken from \citet{HC00}, whereas no \jb\ photometry is available.

In the last two cases, individual extinctions are derived spectroscopically, while the photometric system of the observed NIR photometry is consistent to the 2MASS system at a level of $<$0.1~mag. This uncertainty is lower than the scatter introduced by stellar variability, which typically produces fluctuations of the order of $\gtrsim$0.2~mag \citep{Carpenter2001}.

We correct the observed NIR photometry for interstellar extinction using the extinction law of \citet{Cardelli89} and the values of $A_V$ for individual stars derived spectro-photometrically by D12, R07 and S04. Spectral types are converted into effective temperature \teff\ using the \citet{Luhman03} conversion scale, commonly adopted for PMS stars (Table~\ref{tab:colors}). Our list thus contains 1321 stars with spectral type, luminosity $L$, extinction $A_V$ and extinction-corrected 2MASS-calibrated NIR photometry; 92\% of the sample comes from D10, the remaining part coming from either R07 or S04.

\section{Extraction of the reference stars}\label{sec:analysis}

In order to identify a \lq\lq master\rq\rq\ sample of PMS stars that may more adequately represent the intrinsic colors of the ONC photospheres, we need to clean our sample from field stars, outliers and members with NIR fluxes contaminated by close-in companions, circumstellar emission or strong activity.

\subsection{Age spread}
\begin{figure}
\centering
\includegraphics[width=\linewidth]{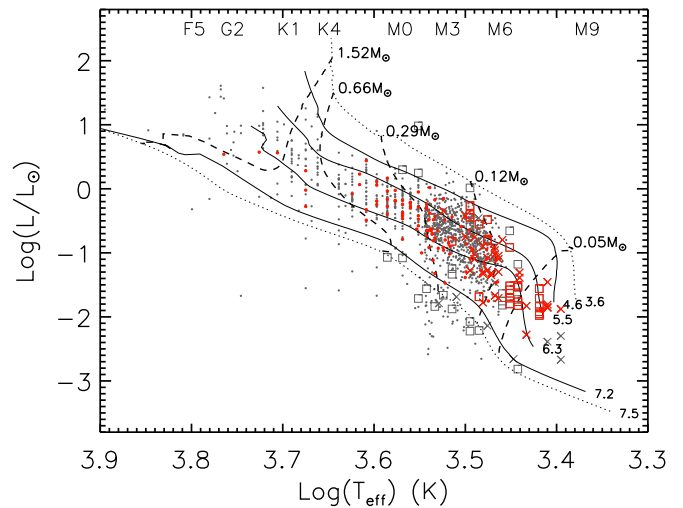}
\caption{HR diagram of our list of stars superimposed to the evolutionary model computed by DM98. Solid lines show a few isochrones at ages included in our selected age range (the logarithmic age in year is indicated), while dashed lines show the tracks at fixed mass. Dots, crosses and squares represent the stars from D10, R07 and S04 respectively. The master sample analyzed in Sect.~\ref{sec:empiric} is shown in red.}\label{fig:hr}
\end{figure}

First, we reject outliers in the HR diagram. Using the list of \teff\ and $L$ compiled in the previous section, we derive ages and masses through a comparison of their positions in the HR diagram with the evolutionary models of DM98 (Fig.~\ref{fig:hr}). The advantage of the DM98 models with respect to the others is that they nicely cover the mass and age ranges spanned by the ONC, while, e.g., the BCAH98 models are available for age$\geq$1~Myr, consistently older than the youngest stars in the HR diagram shown in Fig.~\ref{fig:hr}.

\begin{figure}
\centering
\includegraphics[width=\linewidth]{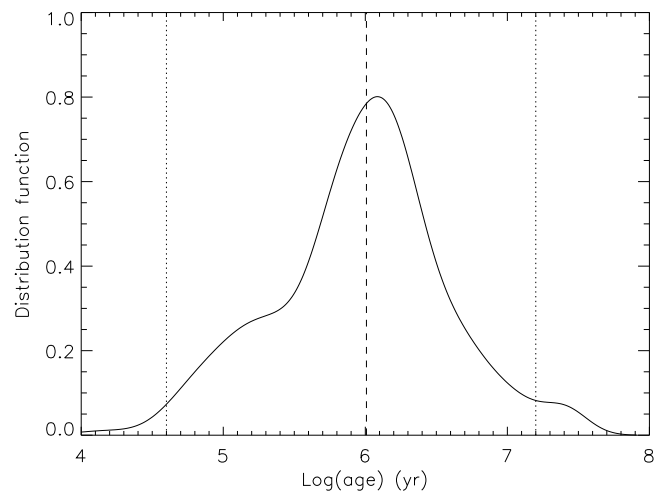}
\caption{Age distribution of the stars based on the DM98 evolutionary models. This plot indicates that the ONC is $\sim$1-2~Myr old. The two dotted lines outline the age range spanned by the analyzed sample of stars.}\label{fig:agespread}
\end{figure}

The distribution of derived logarithmic ages, plotted in Fig.~\ref{fig:agespread}, indicates that according to this model the characteristic age of the ONC members is $\sim$1-2~Myr. Aiming at a large statistical sample, representative of the whole cluster, we conservatively retain all stars with age within the full width at one tenth of the maximum ($4.6<Log(age)<7.2$), rejecting all those that appear too young or too old with respect to the average population.

This selection thus rejects stars too bright compared to the average cluster, i.e.\ stars in the foreground, intrinsically too young or possibly associated to circumstellar excess or close companions.

This selection also discards the stars too faint compared to the average cluster, likely belonging to older stellar associations in the Orion complex \citep{Brown94} or to the galactic background population \citep{Dario11}.

Using other models would have provided different values for individual ages, mean cluster age and lower/upper age limits, but our criteria would have basically selected the same sample. 

\subsection{Stellar multiplicity}

\citet{Tobin2009} find that the frequency of spectroscopic binaries in the ONC is $\sim$10\%, while \citet{Kohler2006} and \citet{Reipurth2007} indicate that the frequency of visual binaries in the cluster is $\lesssim$ 10\% at separations $\gtrsim$60~AU. In particular, \citet{Kohler2006} find that the frequency decreases down to $\sim$5\% across the whole cluster for low-mass systems.

To clean our sample from unresolved multiple systems, we checked the HST/ACS images taken for the Orion Treasury Program \citep{paperone}. This check allows us to discard 40 stars with projected companions at distances $\geq$50~mas (the angular resolution of the images), unresolved in the ISPI images.

For angular distances $\leq$50~mas, corresponding to $\sim$20~AU at the distance of the ONC, any multiple systems would remain unresolved. We thus expect that the fraction of unresolved multiple low-mass systems in our sample is $\lesssim$10\%, as indicated by \citet{Tobin2009}, and does not influence significantly our results.

\subsection{Extinction}\label{sec:extinction_cut}
The extinction estimates in our master sample are derived either photometrically (D10) or spectroscopically (S04, R07). In both cases, the observations are compared to either theoretical or empirical templates. This approach is somehow inaccurate as the effects of surface gravity becomes relevant at spectral types later then $\sim$M3. In order to avoid these uncertainties, we retain the ONC members poorly embedded in the nebula, selecting the stars in our master sample characterized by low extinctions ($A_V<$0.5).

\subsection{Circumstellar Activity}
As shown by \citet{Carpenter2002}, accreting stars are characterized by variability up to 0.2~mags in the NIR. Thus, to avoid the uncertainties introduced by accretion, we reject those stars with significant circumstellar activity.

\begin{figure}
   \centering
   \includegraphics[width=\linewidth]{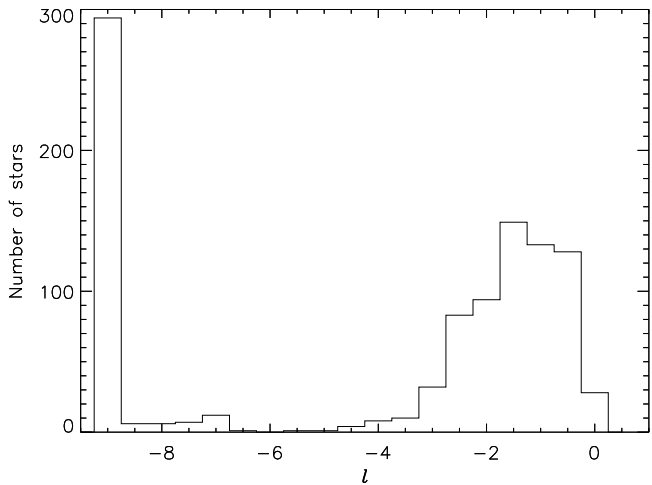}
   \caption{Distribution of the accretion luminosities $l$ in our selected sample. Values $l<$-6 must be regarded as null detection of ongoing accretion, and are kept in the analyzed sample. On the contrary, stars with $l>$-6 show accretion evidences and are discarded, in order to avoid any contamination to the extinction-corrected photometry.}\label{fig:acc_distr}
\end{figure}

For the subsample of stars with $BVI$ photometry taken from \citet{Dario09}, D10 provide indications on ongoing accretion through the parameter $l$ defined as:
\begin{displaymath}
l=Log\left(\frac{L_{accr}}{L_{tot}}\right),
\end{displaymath}
where $L_{tot}$ is the total luminosity of the system and $L_{accr}$ is the accretion luminosity. We discard stars with $l>$-6 (Fig.~\ref{fig:acc_distr}), in order to isolate only those showing no evidence of accretion.

We also exclude the stars in the R07 and S04 sublists with spectroscopic evidence of disk activity, mainly CaII emission lines which are commonly attributed to the presence of disk winds.

\subsection{NIR excess from the inner disk}
\begin{figure}
   \centering
   \includegraphics[width=\linewidth,viewport=1 440 560 845,clip]{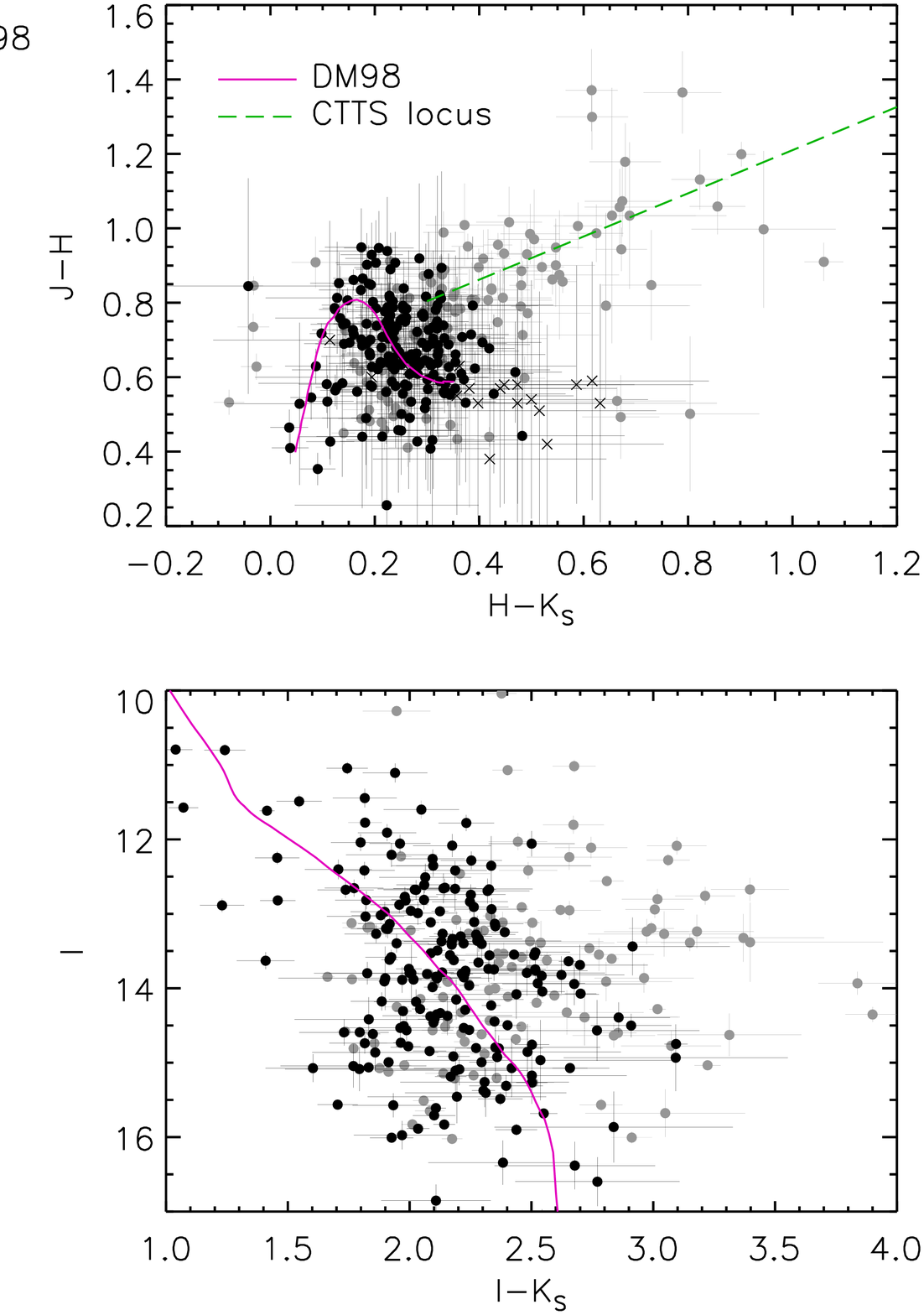}
   \caption{2CD of our sample of stars, compared to the 1~Myr old model provided by DM98 and computed in the 2MASS system (pink solid line). The observed colors do not strictly follow the theoretical model, indeed they occupy a region elongated along the CTTS locus provided by \citet{Meyer97} (green dashed line). Circles and crosses represent stars with spectral types provided by \citet{Dario11} and R07 respectively. Gray error bars represent stars with observed colors not compatible with the model at the 2$\sigma$ level: they are rejected in our analysis.}\label{fig:ccd}
\end{figure}

Circumstellar disks around young PMS stars are known to radiate in the NIR domain \citep[][and \citet{Hillenbrand98} for the specific case of the ONC]{Meyer97,Cieza05,Fischer11}, contaminating the NIR continuum emission of the central stars. To reject sources with photometry contaminated by strong disk excess, we plot in Fig.~\ref{fig:ccd} the NIR 2CD of our 395 stars, comparing their extinction-corrected colors to the 1~Myr isochrone of DM98. 

Whereas the majority of stars appear scattered around the DM98 isochrone, the redder sources (top-right corner of the plot) appear systematically shifted in the region corresponding to the Classical T-Tauri Stars (CTTS) locus proposed by \citet{Meyer97}. These dereddened colors cannot be accounted for by either inaccuracies in the theoretical 2CD or by measurement errors, and must therefore be attributed to circumstellar disk emission. Taking the theoretical isochrone as a proxy for the photospheric colors of the selected stars, we discard the stars with extinction-corrected photometry deviating from the isochrone for more than twice their photometric uncertainties, and we keep the other stars, whose infrared photometry is consistent with the theoretical models within the uncertainties.

To check the robustness of this selection criterion, we cross-match our full NIR catalog \citep{paperI} with the YSOVAR catalog \citep{Morales2011}, containing Spitzer/IRAC photometry of the ONC stars. Based on the observed mid-infrared colors, this catalog also classifies the stars in Class II stars (i.e.\ PMS stars with circumstellar disks) and Class III stars (i.e.\ PMS stars with debris disks, weakly radiating in the NIR).

We analyze the cross-matched catalog by means of the NIR 2CD, as shown above, and we find that $>$60\% of Class II stars are consistently displaced in the diagram along the CTTS locus, while $>$70\% of Class III stars have NIR colors compatible with the theoretical isochrone. This indicates that the majority of the selected stars are Class III stars, while the Class II stars in the retained sample likely have negligible NIR excess, compared to the photometric uncertainties.

This is the last step of our selection process. We end up with a master sample made of 150 bona-fide cluster members with 2500~K$\lesssim$\teff$\lesssim$5000~K, poorly extincted and without substantial evidence of non-photospheric activity, either accretion or circumstellar disk emission.

\section{Accuracy of the \citet{Allard10} atmospheric model}\label{sec:atm}

In this section, we take advantage of the list of stars assembled in Sect.~\ref{sec:dataset} to investigate the accuracy of the synthetic atmospheric templates provided by \citet{Allard10}, which nicely cover the parameters space spanned by our sample of stars.

The assignment of spectral types of late-type young stars is usually performed in the optical-red domain, as this wavelength range provides the best-studied sets of spectral lines for classification and is least affected by veiling from accretion and/or circumstellar emission. With spectral types at hand, it is then possible to obtain a prediction of the photospheric fluxes in a certain passband by convolving the corresponding filter profile with an appropriate template spectrum, once \teff, the logarithmic surface gravity $\log g$ and the metallicity are known. 

For each star in our master sample we already have the effective temperature \teff\ at hand, and we assume the standard solar metallicity for the ONC, as indicated by, e.g., \citet{DOrazi2009} and \citet{Biazzo2011}. In order to compute synthetic photometry, we thus need an estimate of $\log g$. Having the luminosity $L$ and \teff, we first compute the stellar radius $R_*$ using the Stefan-Boltzmann's law $L=4\pi R_*^2\sigma_{SB}T_{\rm eff}^4$. We also derive stellar masses $M_*$ placing the stars in the HR diagram and comparing their position with the theoretical model of DM98 (Fig.~\ref{fig:hr}). We refine the mass estimates taking into account the results of \citet{Hillenbrand2004}, who analyzed the discrepancies between track-predicted masses and dynamical masses for a sample of PMS stars in binary systems. They found that the DM98 models tend to underestimate masses by $\sim$20\%. We then increase by 30\% our track-predicted mass estimates, and we use the resulting values to compute the surface gravities $g$ following the relation $g=GM_*/R_*^2$. 

For each star, we then compute synthetic 2MASS NIR photometry using the atmospheric models of \citet{Allard10}, with the appropriate values for the effective temperature and surface gravity, and the distance d=414$\pm$7~pc\footnote{The uncertainty on the distance of the ONC negligibly propagates onto the uncertainties on the observed photometry.} as derived by \citet{Menten07}. The synthetic magnitudes and colors  $X_{synthetic}$ are then compared to the corresponding extinction-corrected values $X_{extinction-corrected}$, computed taking into account the $A_V$ estimates from optical spectro-photometry (see Sect.~\ref{sec:dataset}). The differences
\begin{equation}
\Delta X(T_{\rm eff})=X_{extinction-corrected}(T_{\rm eff})-X_{synthetic}(T_{\rm eff})\label{eq:deviations}
\end{equation}
represent our proposed corrections to the synthetic NIR colors. These are values to be added to the synthetic colors to derive the correct intrinsic colors of the stars.

\begin{figure}
   \centering
   \includegraphics[width=\linewidth,viewport=15 20 560 730,clip]{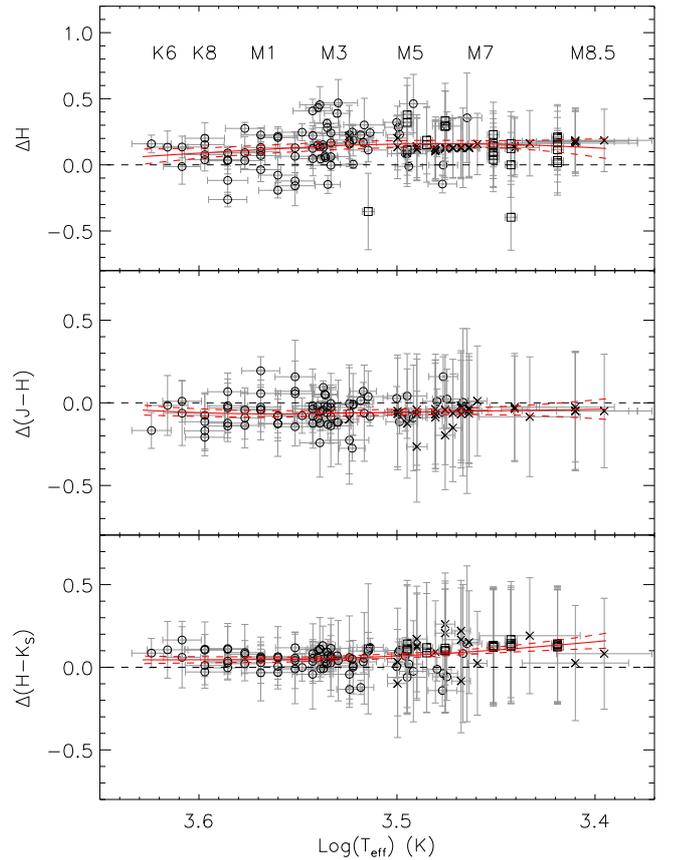}
   \caption{$\Delta H$, $\Delta(J-H)$ and $\Delta(H-K_S)$ corrections to the synthetic colors (from top to bottom respectively). Circles, crosses and squares represent stars with spectral types provided by \citet{Dario11}, R07 and S04 respectively, while the error bars take into account the uncertainties in both the extinction-corrected photometry and the synthetic photometry. Red solid lines show our smooth outlier-resistant fits, together with the corresponding 95\% confidence bands (dashed red lines).}\label{fig:deviations}
\end{figure}

The distribution of stars in terms of \teff\ allows to perform a smooth non-parametric fit of the $\Delta X$ from \teff$\sim$2500~K up to \teff$\sim$4200~K, while the number of stars at higher temperatures is too low to allow for a robust fit (see Fig.~\ref{fig:hr}). The fitted $\Delta H$, $\Delta(J-H)$ and $\Delta(H-K_S)$ for the entire master sample are shown in Fig.~\ref{fig:deviations} and reported in Table~\ref{tab:atm}, as a function of the effective temperature. They indicate that the \hb\ magnitude is underestimated by the model (i.e. the model systematically provides brighter \hb\ magnitudes) by $\lesssim$0.2 magnitudes, with an uncertainty $\lesssim$0.1~magnitude. The same is true for the colors, with a smaller offset of the order of $\lesssim$0.1 magnitudes, in the red and blue directions for the \jb-\hb\ and \hb-\kb\ colors respectively.

\begin{table}
  \centering
  \caption{Corrections to synthetic colors for LMSs and BDs in the ONC.}\label{tab:atm}
  \begin{tabular}{ll|ccc}
    \hline\hline
    SpT & \teff\tablefootmark{a} & $\Delta$\hb & $\Delta$(\jb-\hb) & $\Delta$(\hb-\kb)\\
    \hline
M8.5 & 2555 &    0.13 &   -0.04 &    0.15\\
M8 & 2710 &    0.15 &   -0.05 &    0.12\\
M7 & 2880 &    0.16 &   -0.05 &    0.10\\
M6 & 2990 &    0.16 &   -0.05 &    0.08\\
M5 & 3125 &    0.16 &   -0.06 &    0.07\\
M4 & 3270 &    0.15 &   -0.06 &    0.06\\
M3 & 3415 &    0.14 &   -0.06 &    0.05\\
M2 & 3560 &    0.13 &   -0.07 &    0.05\\
M1 & 3705 &    0.12 &   -0.07 &    0.05\\
M0 & 3850 &    0.10 &   -0.06 &    0.05\\
K8 & 3965 &    0.09 &   -0.06 &    0.05\\
K7 & 4060 &    0.08 &   -0.06 &    0.05\\
K6 & 4154 &    0.07 &   -0.05 &    0.04\\
    \hline
    \end{tabular}
\tablefoot{
   \tablefoottext{a}{Temperature scale of \citet{Luhman03}.}
}
\end{table}

\begin{figure}
   \centering
   \includegraphics[width=\linewidth,viewport=15 440 560 730,clip]{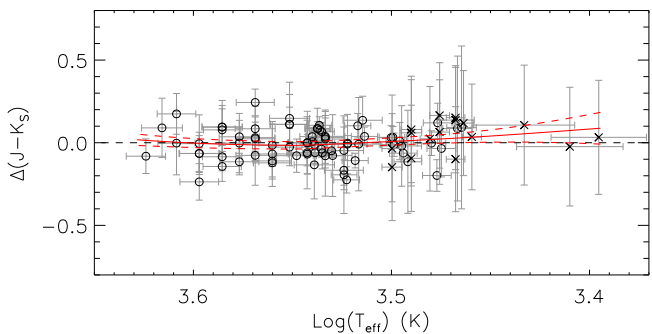}
   \caption{$\Delta(J-K_S)$ corrections to the synthetic colors. Symbols and colors are the same as in Fig.~\ref{fig:deviations}.}\label{fig:deviations_jk}
\end{figure}

\begin{figure*}
   \centering
   \includegraphics[width=\linewidth]{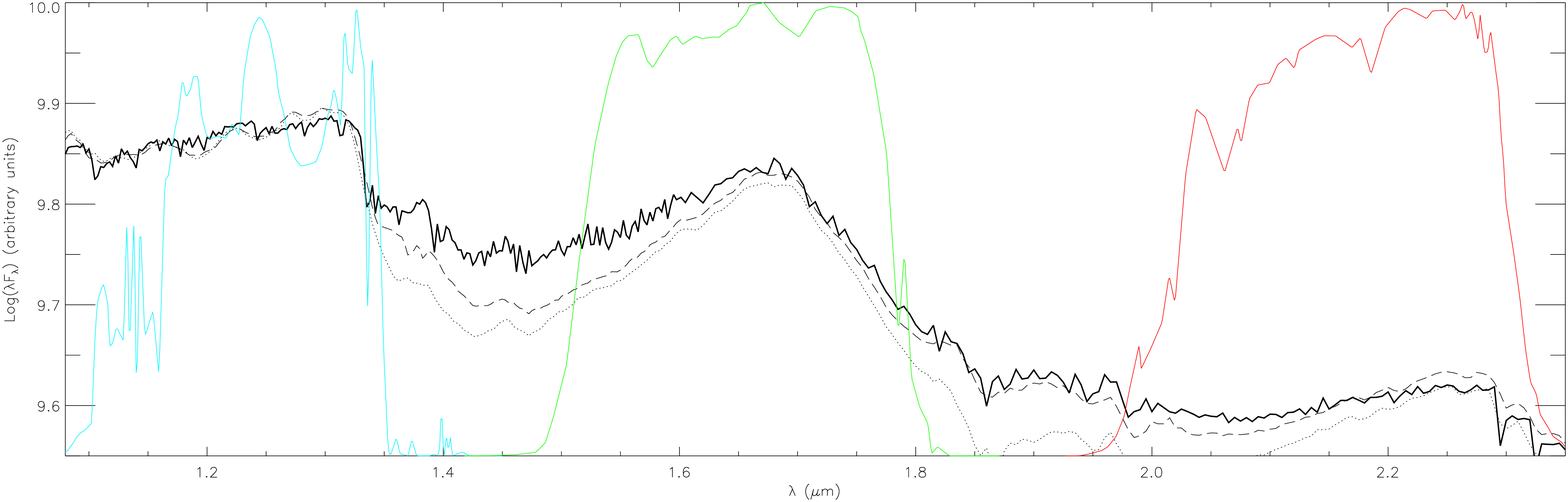}
   \caption{Synthetic spectra of a M7 PMS star (solid line) as provided by \citet{Allard10}, together with the observed spectra of two young M7 stars \citep[in dots and dashes, ][]{Muench07}. The blue, green and red lines stand for the transmission curves of the 2MASS \jb\hb\kb\ filters respectively. We find good agreement between the model and the observations in the \jb\kb\ bands, while in the \hb\ band the model overestimates the radiated flux (or underestimates the opacity).}\label{fig:spectra_emp}
\end{figure*}

One possible explanation for these discrepancies is that the atmospheric model provides \hb\ fluxes larger than the observed ones, while providing fairly good predictions on the \jb\kb\ fluxes. From an observational point of view, this hypothesis is consistent with Fig.~\ref{fig:deviations_jk}, where we show that the atmospheric model generally provides consistent predictions on the \jb-\kb\ colors. In order to address this issue, in Fig.~\ref{fig:spectra_emp} we compare the synthetic M7 spectrum of \citet{Allard10} with the spectra of two M7 stars observed by \citet{Muench07} in IC 348 \citep[age$\sim$2~Myr,][]{Luhman03}. This plot confirms that, despite the improvements in the latest release, the model still does not fully account for the continuum opacity of water vapor, which significantly affects the NIR SED, and in particular the \hb-band continuum, of cool stars.

Another possibility is that the atmospheric models do not account for the collision induced absorption (CIA) from molecular hydrogen H$_2$. Theoretical models computed by \citet{Borysow1997} show that CIA is relevant for the infrared spectrum of cool subgiants: radiative energy is absorbed in the infrared and re-emitted at visible wavelengths, leading to the typical triangular shape of the \hb-band \citep[][ see also Fig.~\ref{fig:spectra_emp} and Fig.~\ref{fig:spectra} in this paper]{Allers2007,Kirkpatrick2008}.

\subsection{Systematic uncertainties}\label{sec:uncertainties_atm}

The corrections we have derived are based on the reddening law used to correct the observed photometry for extinction and on the accuracy of the mass estimates. The results may be influenced by the assumed ratio of total to selective extinction $R_V$ (Sect.~\ref{sec:rv}) and by the accuracy of the mass (and surface gravity) estimates derived from the DM98 theoretical model (Sect.~\ref{sec:masses}). 

\subsubsection{Effects of the assumed $R_V$}\label{sec:rv}

In our derivation of the intrinsic NIR colors of stars in the ONC, we have assumed that the total to selective extinction is $R_V=A_V/E(B-V)$=3.1, which is the typical Galactic value. However, it has been proposed that the reddening law toward the OB stars of the Orion association can be better parametrized by $R_V \simeq 5.5$, typical of larger dust grains \citep{Johnson1967,Costero1970,Baldwin1991,Osterbrock1992,Greve1994,Blagrave2007}.

We thus investigate how our results vary with increasing $R_V$ considering the subsample of stars of D10 (corresponding to $\sim$92\% of our collected catalog down to spectral type M7), for which the authors provide the stellar parameters derived assuming $R_V$=5.5. We do not include the list of stars of R07 and S04 in this analysis as the authors do not derive the stellar parameters assuming $R_V=$5.5. By consequence, we can derive the $\Delta X$ corrections only down to the M7 spectral type.

\begin{figure}
   \centering
   \includegraphics[width=\linewidth,viewport=1 1 540 725,clip]{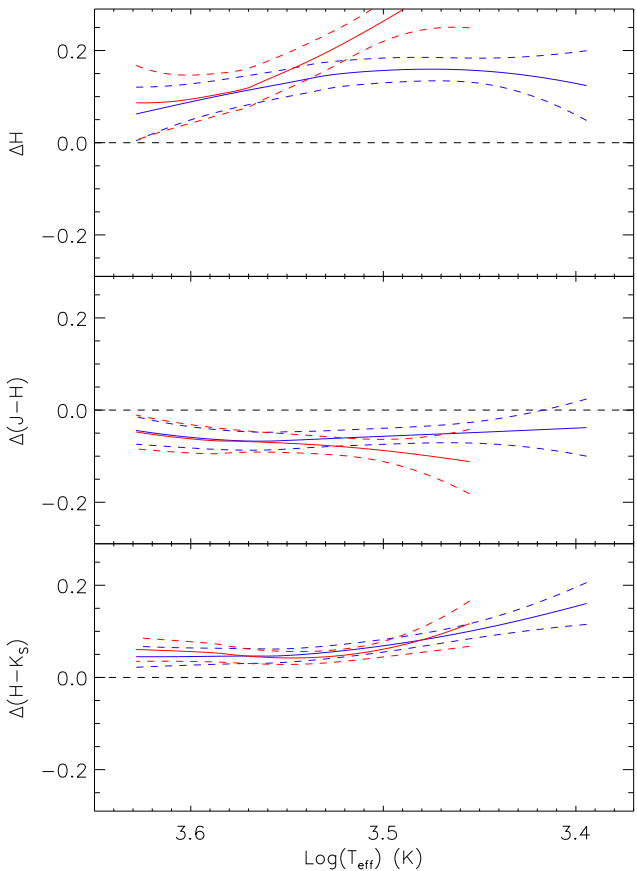}
   \caption{Comparison of the $\Delta H$, $\Delta(J-H)$ and $\Delta(H-K_S)$ corrections (from top to bottom respectively) derived assuming $R_V$=3.1 (blue lines) ans $R_V$=5.5 (red lines). The dashed lines represent the 95\% confidence bands.}\label{fig:deviations_comp}
\end{figure}

The new set is analyzed following the same strategy described in Sect.~\ref{sec:analysis}. Using the new set of the stellar parameters $L$, $A_V$ and $l$, we compile a catalog of 117 stars and derive new corrections to the NIR synthetic colors (Table~\ref{tab:atm}). In Fig.~\ref{fig:deviations_comp} we compare these new corrections to the ones obtained with $R_V$=3.1, finding that the two sets of corrections are consistent within the errors.

Moreover, as shown by \citet{Scandariato10}, the foreground Orion Nebula generally provides magnitudes $A_V\lesssim$2~mag. Thus, the cut at low extinctions discussed in Sect.~\ref{sec:extinction_cut} excludes all those stars deeply embedded in the nebula. We also remark that the same $R_V$  holds for higher extinctions within the Orion Nebula\, as indicated by e.g.\ \citet{Indebetouw2005} and \citet{Dario10}.

For these reasons, and considering the weak dependence of IR extinction on different values of $R_V$, from now on we will adopt for simplicity the case $R_V$=3.1.

\subsubsection{Surface gravity estimate}\label{sec:masses}

Our analysis of the synthetic photometry has been based on a particular family of evolutionary models, DM98, used to derive the stellar mass and therefore the surface gravity. This assumption may affect especially the NIR continuum of M-type stars, shaped by the water vapor opacity profile \citep{Allard10}, which is sensitive to surface gravity. In order to reduce the uncertainties in $\log g$ related to the assumption of a particular model, in Sect.~\ref{sec:atm} we increased the DM98 track-predicted masses by $\sim$30\%, as indicated by \citet{Hillenbrand2004}. Since mass and surface gravity are linearly correlated, this correction increases $\log g$ by $\sim$0.16~dex with respect to the value derived using the DM98 model masses at face value.

\begin{figure*}
\centering
\includegraphics[width=\linewidth]{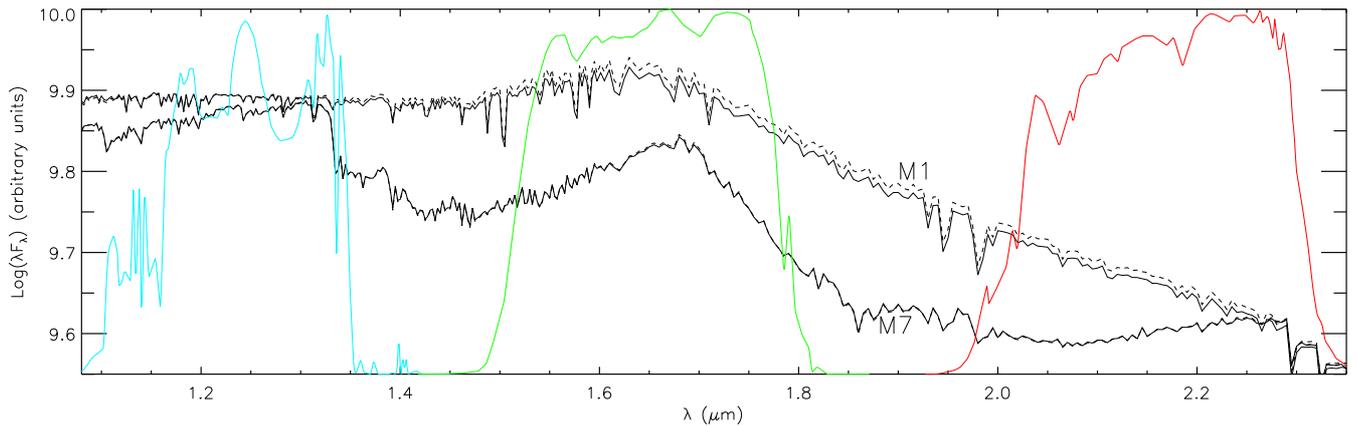}
\caption{Synthetic spectra for a M1 and a M7 star (top and bottom spectrum respectively) as provided by \citet{Allard10}. For each spectral type, the dashed line shows the spectrum computed with $\log g$ provided by the 1~Myr old model of DM98 (3.63 and 3.33 respectively), while the solid line shows the same spectra with $\log g$ artificially increased by 0.16 dex (for the M7 star the two spectra overlap). The blue, green and red lines stand for the transmission curves of the 2MASS \jb\hb\kb\ filters respectively. We find that the M7 spectrum weakly depends on $\log g$. On the contrary, for the M1 star the continuum spectrum in the \hb\kb\ bands decreases with increasing $\log g$. No changes are seen in the \jb\ band.}\label{fig:spectra}
\end{figure*}

To further investigate the effects of surface gravity in our analysis, in Fig.~\ref{fig:spectra} we compare the synthetic near-IR spectra of an M1 and M7 star both 1~Myr old according to DM98. The dashed and solid lines show the spectra before and after the mass correction, respectively. For the M7 star the correction has no impact. On the other hand, the \hb\kb\ continuum of the M1 star decreases by $\lesssim$0.01~dex, while no significant changes are present in the \jb\ band. In terms of synthetic photometry, the increase of surface gravity leads to fainter \hb\kb\ magnitude and bluer \jb-\hb\ colors by $\lesssim$0.03 magnitudes, which is less than the uncertainties on the atmospheric model (see, e.g., Fig.~\ref{fig:spectra_emp}). Thus, fine tuning the $\log g$ values has a negligible effect on the synthetic photometry and, by consequence, the differences between predicted and observed photospheric colors are not dominated by the uncertainties in the evolutionary models, as long as the track-predicted masses deviate by $\lesssim$30\%.

If using different recipes for deriving the stellar mass, and therefore surface gravity, has a negligible effect on the synthetic spectra of PMS stars, the discrepancies we found between model predictions and extinction-corrected photometry (Table~\ref{tab:atm}) must be largely attributed to the residual uncertainties of the atmospheric models of \citet{Allard10} in the NIR. 

\section{The empirical NIR isochrone of the ONC}\label{sec:empiric}

\begin{figure}
   \centering
   \includegraphics[width=\linewidth,viewport=600 1 1120 680,clip]{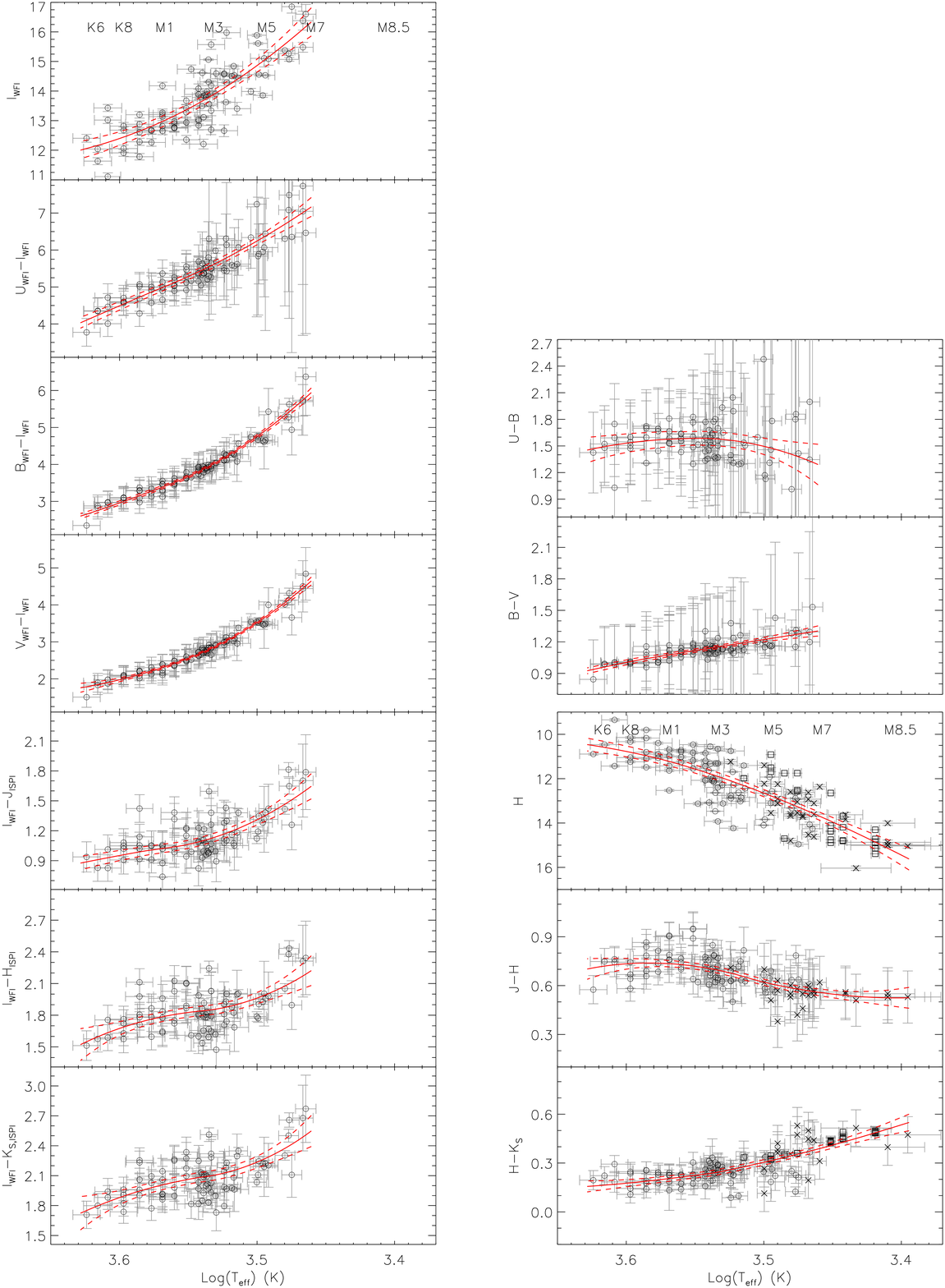}
   \caption{Intrinsic \hb, \jb-\hb\ and \hb-\kb\ colors of stars in the ONC as function of \teff\ (from top to bottom respectively). Circles, crosses and squares represent stars with spectral types provided by \citet{Dario11}, R07 and S04 respectively, while the error bars take into account the uncertainties in both the extinction-corrected photometry and the synthetic photometry. Red solid lines show our fits, together with the corresponding 95\% confidence bands (dashed red lines).}\label{fig:colors}
\end{figure}

By taking advantage of the sample of ONC stars compiled in Sect.~\ref{sec:dataset}, we can empirically derive a representative NIR isochrone of the cluster, assuming $R_V$=3.1. In Fig.~\ref{fig:colors} we show the extinction-corrected \hb, \jb-\hb\ and \hb-\kb\ colors of the sample of stars analyzed in Sect.~\ref{sec:analysis} as functions of the effective temperature. Following the same strategy adopted in Sect.~\ref{sec:atm}, we smoothly fit the intrinsic colors from \teff$\sim$2500~K up to \teff$\sim$4200~K (3.40$\lesssim$Log(\teff)$\lesssim$3.62) using a local polynomial regression smoother. The derived intrinsic NIR colors of the ONC are given in Table~\ref{tab:colors}.

\begin{table}
  \centering
  \caption{Empiric intrinsic colors for LMSs and BDs in the ONC.}\label{tab:colors}
  \begin{tabular}{ll|ccc}
    \hline\hline
    SpT & \teff\tablefootmark{a} & \hb & \jb-\hb & \hb-\kb\\
    \hline
M8.5 & 2555 &   15.21 &    0.52 &    0.52\\
M8 & 2710 &   14.44 &    0.53 &    0.45\\
M7 & 2880 &   13.69 &    0.56 &    0.39 \\
M6 & 2990 &   13.26 &    0.58 &    0.35 \\
M5 & 3125 &   12.78 &    0.61 &    0.31 \\
M4 & 3270 &   12.31 &    0.66 &    0.28 \\
M3 & 3415 &   11.90 &    0.70 &    0.24 \\
M2 & 3560 &   11.54 &    0.72 &    0.22 \\
M1 & 3705 &   11.22 &    0.73 &    0.20 \\
M0 & 3850 &   10.96 &    0.74 &    0.19 \\
K8 & 3965 &   10.79 &    0.73 &    0.18 \\
K7 & 4060 &   10.67 &    0.73 &    0.17 \\
K6 & 4154 &   10.56 &    0.72 &    0.16 \\
    \hline
    \end{tabular}
\tablefoot{
   \tablefoottext{a}{Temperature scale of \citet{Luhman03}.}
}
\end{table}

The magnitudes and colors presented in Table~\ref{tab:colors} represent our best estimate of the locus in the NIR CMDs along which the ONC stars should be located if one removes the effects of extinction and NIR excess, and there is no intrinsic color dispersion among stars. We underline that this is an average, ideal locus. The ONC population is spread across the HR diagram (see e.g.  Fig.~\ref{fig:hr}) by about $\sim$1.5~dex in luminosity, partially due to effects like stellar variability and unresolved companions. Moreover, recent studies indicate that the luminosity spread may be ascribed to an intrinsic age spread \citep{Hillenbrand2009,Reggiani11}. The observed luminosity spread, $\sim$3.5 magnitude, is reflected in the top panel of Fig.~\ref{fig:colors} by a spread along the vertical axis larger than the error bars. Our empirical \hb\ magnitudes, therefore, may only be appropriate for an average ONC isochrone.

On the other hand, we have seen that the differences in luminosity, while leading to differences in surface gravity, have small influence on the photospheric colors (Sect.~\ref{sec:masses}). For this reason the average NIR colors in Fig.~\ref{fig:colors} are better constrained than the absolute magnitudes. One can therefore assume that our empirical, average {\sl colors} properly describe those of the ONC sources, and of their corresponding isochronal sequences.

\begin{figure}
   \centering
   \includegraphics[width=\linewidth]{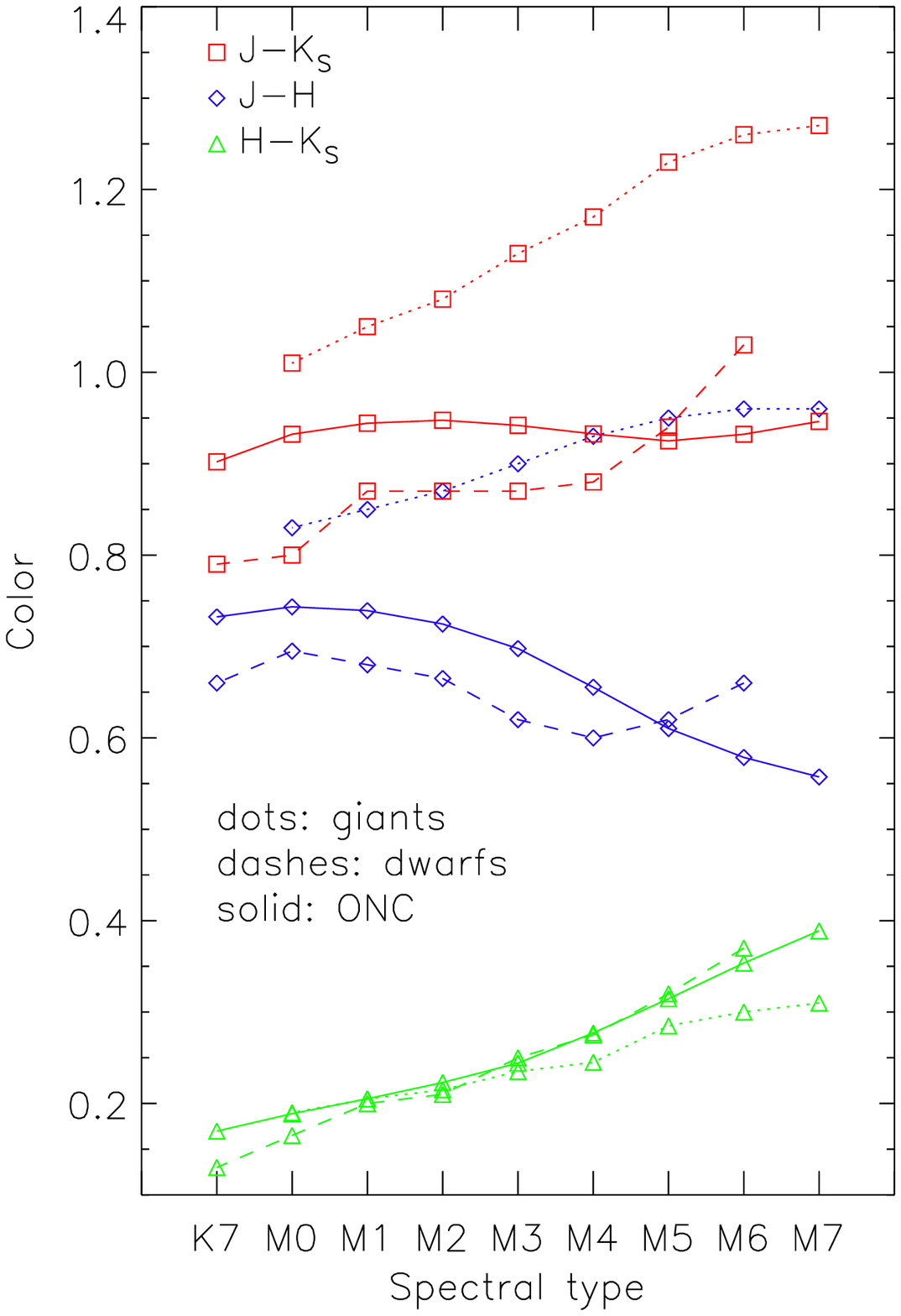}
   \caption{\jb-\hb\ (blue diamonds), \hb-\kb\ (red triangles) and \jb-\kb\ (green squares) colors of the ONC (solid line), giants (dotted lines) and dwarfs (dashed lines), as provided by \citet{Bessel1988}.}\label{fig:bessel}
\end{figure}

In Fig.~\ref{fig:bessel} we compare our empirically-derived colors for the ONC stars with the corresponding observed average colors of giants and dwarfs provided by \citet{Bessel1988}. This plot shows that the \jb-\hb\ and \jb-\kb\ colors generally get bluer with increasing gravity, while the \hb-\kb\ color is less affected. This is consistent with Fig.~\ref{fig:spectra}, where we show that the \hb\ and \kb\ fluxes tend to increase with $\log g$, while the \jb\ flux is not sensitive to surface gravity. Moreover, we remark that the ONC stars, being PMS stars, have intermediate gravities between giants and dwarfs, and this is reflected by the fact that they have NIR colors intermediate between the giants' and dwarfs' colors.

\subsection{Test of methodology}
As a sanity check, we now test our methodology against a sample of main sequence stars in the range 5,000~K$\leq$\teff$\leq$6,000~K (corresponding to the K1--G0 spectral type in the temperature scale of \citet{SK1982}), where theoretical models are known to produce robust predictions on photospheric colors. We select 77 stars in the PASTEL catalog \citep{Soubiran2010}, with solar metallicities, surface gravities and temperatures derived spectroscopically, and \jb\hb\kb\ magnitudes extracted from the 2MASS catalog. These stars are located within 30~pc from the Sun and, by consequence, the observed NIR photometry is negligibly affected by interstellar extinction.

\begin{figure}
   \centering
   \includegraphics[width=\linewidth,viewport=590 1 1120 680,clip]{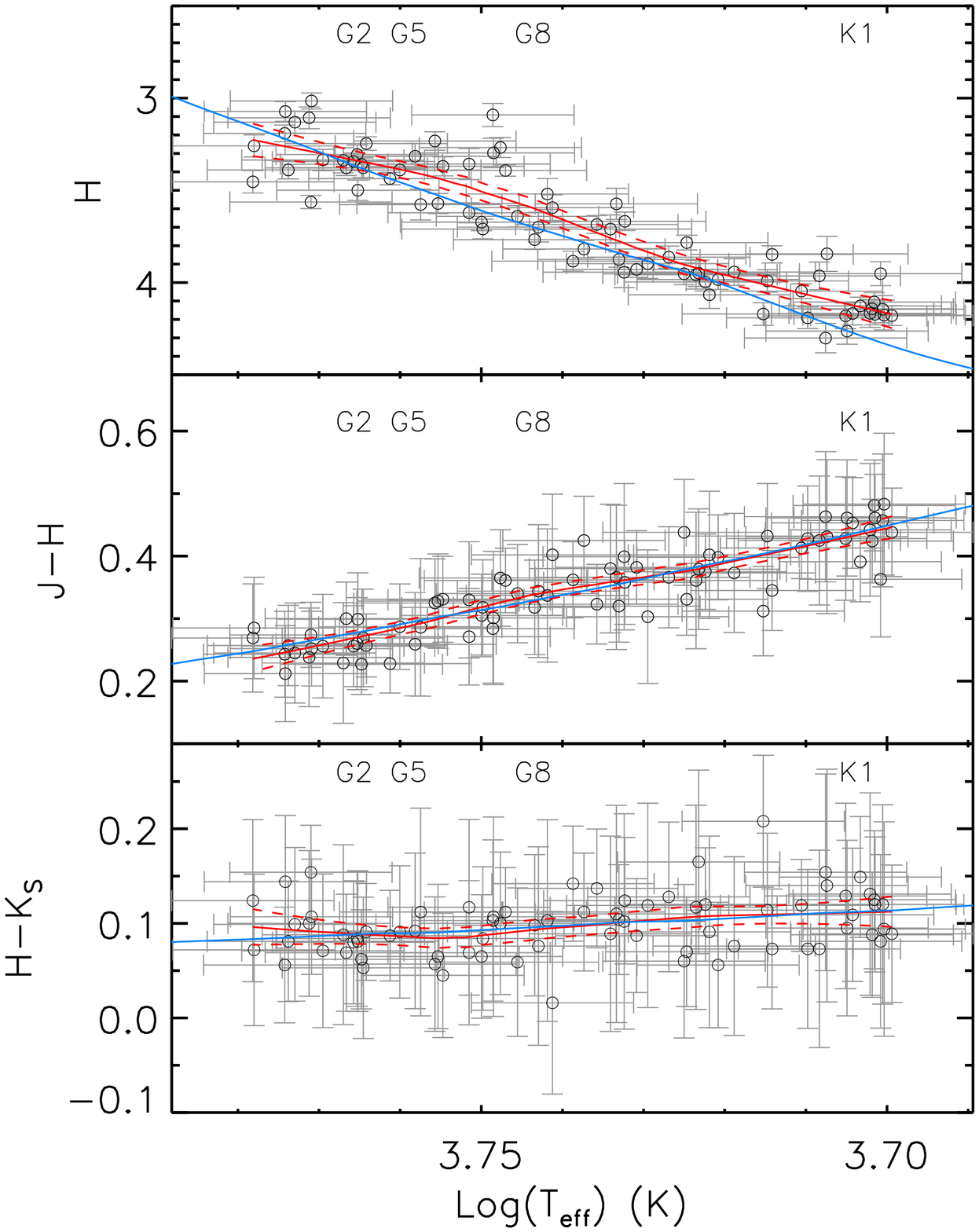}
   \caption{Observed \hb, \jb-\hb\ and \hb-\kb\ colors of main sequence stars in the in the solar neighborhood as function of \teff\ (from top to bottom respectively). Red solid lines show our fits, together with the corresponding 95\% confidence bands (dashed red lines), while the cyan lines represent the synthetic colors computed as described in the text.}\label{fig:ms_isoch}
\end{figure}

In Fig.~\ref{fig:ms_isoch} we plot our best fits of the photospheric colors of the selected stars as functions of \teff, obtained as described in Sect.~\ref{sec:atm}. We also compare our empirical fits with the synthetic photometry computed combining the evolutionary model of \citet{Dantona98} with the atmospheric model of \citet{Allard10}. These models have been extensively tested against main sequence stars, and have proved to consistently reproduce the observed NIR magnitudes and colors for the analyzed spectral range within the photometric uncertainties. A posteriori, we fix the inaccuracies of the atmospheric model described in Sect.~\ref{sec:atm}, adding 0.05 magnitudes to the synthetic \hb\ magnitudes, thus extrapolating Table~\ref{tab:atm} towards earlier spectral types.

Figure~\ref{fig:ms_isoch} shows that our algorithm recovers the expected magnitudes and colors within the 95\% confidence bands. This test thus confirms the consistency of our analysis, providing additional confidence to the empirical colors listed in Table~\ref{tab:colors}.

\section{Comparison with theoretical models}\label{sec:discussion}

\subsection{Magnitudes and colors vs. \teff}
As anticipated in Sect.~\ref{sec:intro}, current evolutionary models provide different predictions for the position of the PMS isochrones and evolutionary tracks in the HR diagram, especially at sub-solar masses and young ages  \citep{Hillenbrand2008}. This is a consequence of the difficulties in adequately modeling the convective equilibrium in the interior of fast rotating sub-solar young stars.

Alongside with the differences between the adopted input parameters and theoretical treatment, theoretical models also differ in setting the $t=0$ instant of the PMS evolution, i.e.\ there is no absolute calibration of the \lq\lq clock\rq\rq\ counting the age of young stars. For this reason, theoretical models may also mismatch. 

The  inconsistencies between the various sets of models generally persist when they are converted into magnitudes and colors, for a direct comparison with the data in the CMDs  (see Fig.~\ref{fig:isocs3.1}). As discussed in Sect.~\ref{sec:masses}, the synthetic magnitudes depend weakly on the exact value of the mass associated to a given \teff\ and therefore these discrepancies are almost entirely a consequence of the different luminosities predicted by theoretical model, for any particular \teff. 

To illustrate this point, we have computed synthetic photometry for the 1~Myr theoretical isochrones of DM98, and the 2~Myr isochrones of SDF00, which are adequate to represent the typical ONC population (see D10). The different ages of these models are likely due to different age scales, as discussed above.

We also consider the 2~Myr old DUSTY model \citep{Chabrier2000} of the Lyon group, well suited for cool atmospheres (\teff$\lesssim$2900~K). In order to cover the spectral range of our selected sample of stars, we extend the DUSTY model to hotter temperatures using the BCAH98 model with mixing length parameter $\alpha$=1.5, which, a posteriori, provides the best match between theoretical predictions and our empirical isochrone.

\begin{figure*}
\centering
\includegraphics[viewport=1 1 566 725,clip,width=.5\linewidth]{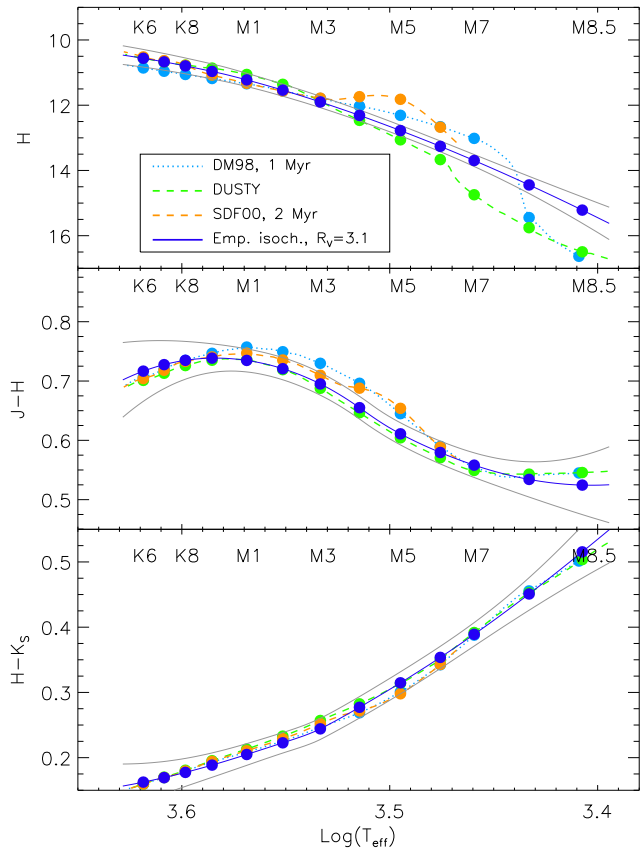}
\caption{\hb, (\jb-\hb) and (\hb-\kb) as functions of \teff\ (from top to bottom). At the top of each panel we report the spectral type--\teff\ conversion of  \citet{Luhman03}. In each panel, dots represent the spectral types listed in Table~\ref{tab:colors} for which the correction is available (see Table~\ref{tab:atm}). The thin gray solid lines represent the uncertainties on our derived empirical colors, as shown in Fig.~\ref{fig:colors}.}\label{fig:cm_temp}
\end{figure*}

We used the synthetic spectra of \citet{Allard10} and the empirical corrections to the atmospheric model derived in Sect.~\ref{sec:atm}. In Fig.~\ref{fig:cm_temp} we plot \hb, (\jb-\hb) and (\hb-\kb) as functions of \teff\ for the various models.

Figure~\ref{fig:cm_temp} shows that there are significant differences in the \hb--\teff\ relation. The DUSTY curve provides the closest match to our empirical model down to spectral type M5; for later types, it drastically turns down such to overpredict $\log$\teff\ by $\sim$0.4 (corresponding to $\sim$250~K). The same discrepancy is found by \citet{Dupuy2010}, who compared the DUSTY models with the NIR spectra of late-M dwarfs with high-quality dynamical masses. Based on their findings, they suggest that the atmospheric models are more likely than the evolutionary models to be the primary source of the discrepancy, since roughly the same \teff\ offset is observed over a wide range of masses, ages, and activity levels but the same temperature range.

Small differences are present between the DUSTY and SDF00 isochrones in the overlapping spectral range, with the noticeable exception that the latter shows a bump in luminosity for M5 stars. On the other hand, the DM98 model is slightly flatter: it overestimates the temperature of late K stars and underestimates that of mid-M types. Then, at very late spectral types, it steeply decreases reconciling with the BCHA98 model. This because the DM98 isochrones drastically turn down in luminosity for $\log$\teff$\lesssim$3.45, as shown in Fig.~\ref{fig:hr}. As discussed by \cite{Hillenbrand2008}, this drop is a common feature of the PMS evolutionary models, which tend to under-predict the intrinsic luminosities of low-mass stars.

Figure~\ref{fig:cm_temp} also shows that despite the discrepancies between the various \hb--\teff\ relations, the situation is much better in what concerns synthetic colors. In this case, model predictions are consistent within $\sim$0.03~mag. In particular, the synthetic \jb-\hb\ colors of the DM98 and SDF00 models are redder by $\sim$0.03~mag than our empirical isochrone. This is due to the fact that, for early to mid M types, the masses provided by these models are $\sim$40\% lower than the corresponding masses given by the DUSTY model, and this leads to the color shift discussed in Sect.~\ref{sec:masses}. This confirms that the discrepancies are mainly due to different predictions of the stellar masses, which reflects in the predicted \hb\ magnitudes.

Our empirical \hb, (\jb-\hb) and (\hb-\kb) are also shown in Fig.~\ref{fig:cm_temp} as functions of \teff. The \hb--\teff\ curve provides higher fluxes for spectral types $\geq$M7 than the DUSTY and DM98 models. In general, no model provides \hb\ magnitudes nicely consistent with our empirical curve. On the other hand, our empirical colors are fully consistent, within the error bars, with the model predictions.

\subsection{Color-Magnitude Diagrams}

\begin{figure*}
\centering
\includegraphics[width=.8\linewidth,viewport=1 1 567 425, clip]{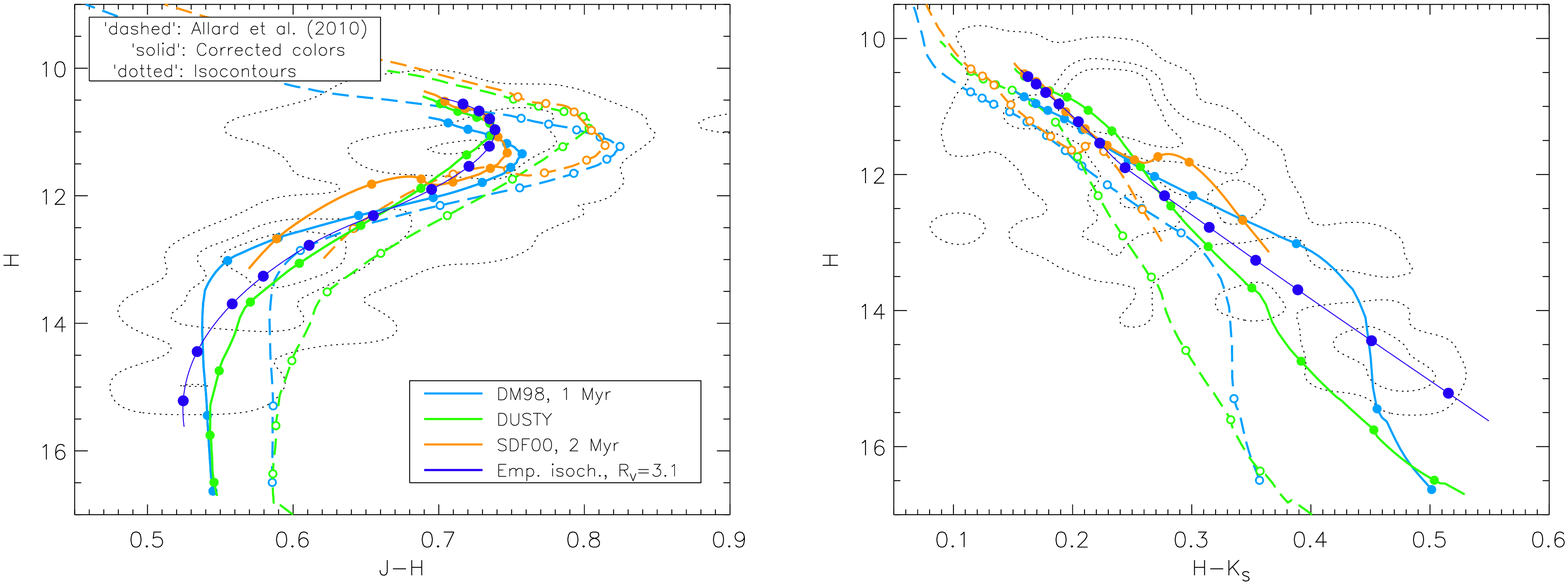}
\includegraphics[width=.8\linewidth,viewport=568 1 1134 425, clip]{cm3.1.eps}

\caption{(\hb,\jb-\hb) and (\hb,\hb-\kb) CMDs (top and bottom panel respectively) of the ONC, compared to the 1~Myr old isochrone model of DM98 (cyan line), the 2~Myr old isochrone of \citet{Chabrier2000} (green line) and the 2~Myr old isochrone of \citet{Siess00} (green line). Synthetic photometry for these models has been computed using the grid of atmospheres provided by \citet{Allard10} (in dashes) and then corrected using the corrections in Table~\ref{tab:atm} (solid lines), assuming $R_V$=3.1. Dots mark the position of the spectral types listed in Table~\ref{tab:colors} along the isochrones (from K6 to M8.5 with increasing magnitude, from K6 to M6 for the model of SDF00). The dotted lines show the iso-density contours for the analyzed stellar sample.}\label{fig:isocs3.1}
\end{figure*}

Figure~\ref{fig:isocs3.1} summarizes these relations in the CMDs, together with the density distribution of the extinction-corrected photometry of our sample of stars. The agreement between the various isochrones is generally poor, as the predicted magnitudes may differ by up to 1~mag.

The synthetic colors corrected for the inaccuracies of the atmospheric models (Sect.~\ref{sec:atm}, solid lines in Fig.~\ref{fig:isocs3.1}) improve over the original theoretical predictions (dashed lines), tracing more closely the color distribution of the ONC sources in the CMDs. This correction, however, does not fix the systematic under-prediction of absolute magnitudes. Our empirical isochrone, derived from a subset of these data, also traces the color distributions of the sources in both CMDs and provides brighter \hb\ magnitudes for late type stars.

\section{The accuracy of the photometrically-determined IMF of the ONC}\label{sec:imf}

The discrepancies between the intrinsic magnitudes predicted by different evolutionary models affect the derivation of the main 
physical parameters of individual stars and clusters. In this section we analyze their impact on the reconstruction of the IMF of a young cluster like the ONC and the possible improvement attained by using our empirical magnitudes and colors. To this purpose, we run an artificial experiment assuming a mass distribution of the ONC and our empirical isochrone to produce \lq\lq true\rq\rq\ NIR magnitudes. Then, we recover the luminosity function (LF) of the cluster and the IMF by using the various model relations.

Let us remind first that in order to analyze the IMF of a cluster like the ONC (for simplicity, we shall assume that the mass distribution of a young cluster still reflects the IMF), a common approach is to collect NIR photometric data to derive, after some assumptions on the age, distance and reddening, the intrinsic LF $\frac{dN}{dm}$ of the population in a generic photometric band $m$. The next step is to convert the LF of the sample into the underlying mass distribution using an appropriate Mass-Luminosity Relation (MLR), $m(M)$ \citep[see e.g.][]{Dantona98,Lada03}. The mass distribution $\frac{dN}{dM}$ can then be derived as
\begin{equation}
\frac{dN}{dM}(M)=\frac{dN}{dm}(m(M))\cdot\frac{dm}{dM}(M).\label{eq:imf}
\end{equation}
In most cases, the MLR is based on theoretical predictions.

In order to simulate a statistically significant stellar sample, we randomly generate 10$^5$ stars with mass distributed according to the multi-part power-law mass function derived by \citet{Muench02} for the ONC (Fig.~\ref{fig:imf}), and extending the trend at very low masses down to $M\sim0.02M_\odot$ (thus neglecting the secondary peak they report at sub-stellar masses).

We then convert masses into effective temperatures, and then into photospheric \jb\ magnitudes. We focus on the \jb-magnitudes because this band is usually assumed to be the most suitable one (together with the $I$ and \hb\ bands) in deriving the IMF of young clusters, being relatively unaffected by accretion, disk emission and other circumstellar activity \citep[see e.g.\ ][]{Hillenbrand97,LR00}.

\begin{figure}
\centering
\includegraphics[width=\linewidth]{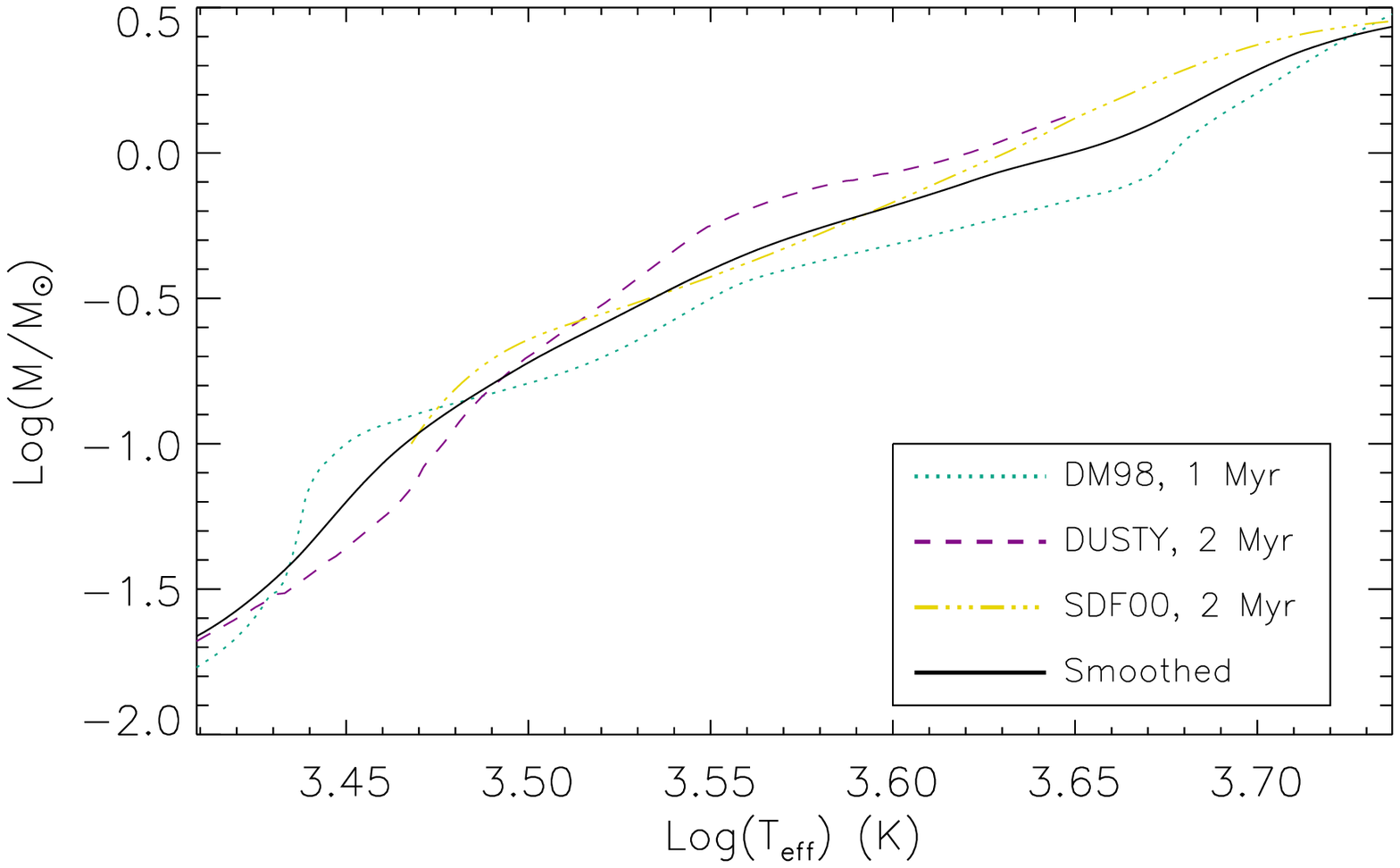}
\includegraphics[width=\linewidth]{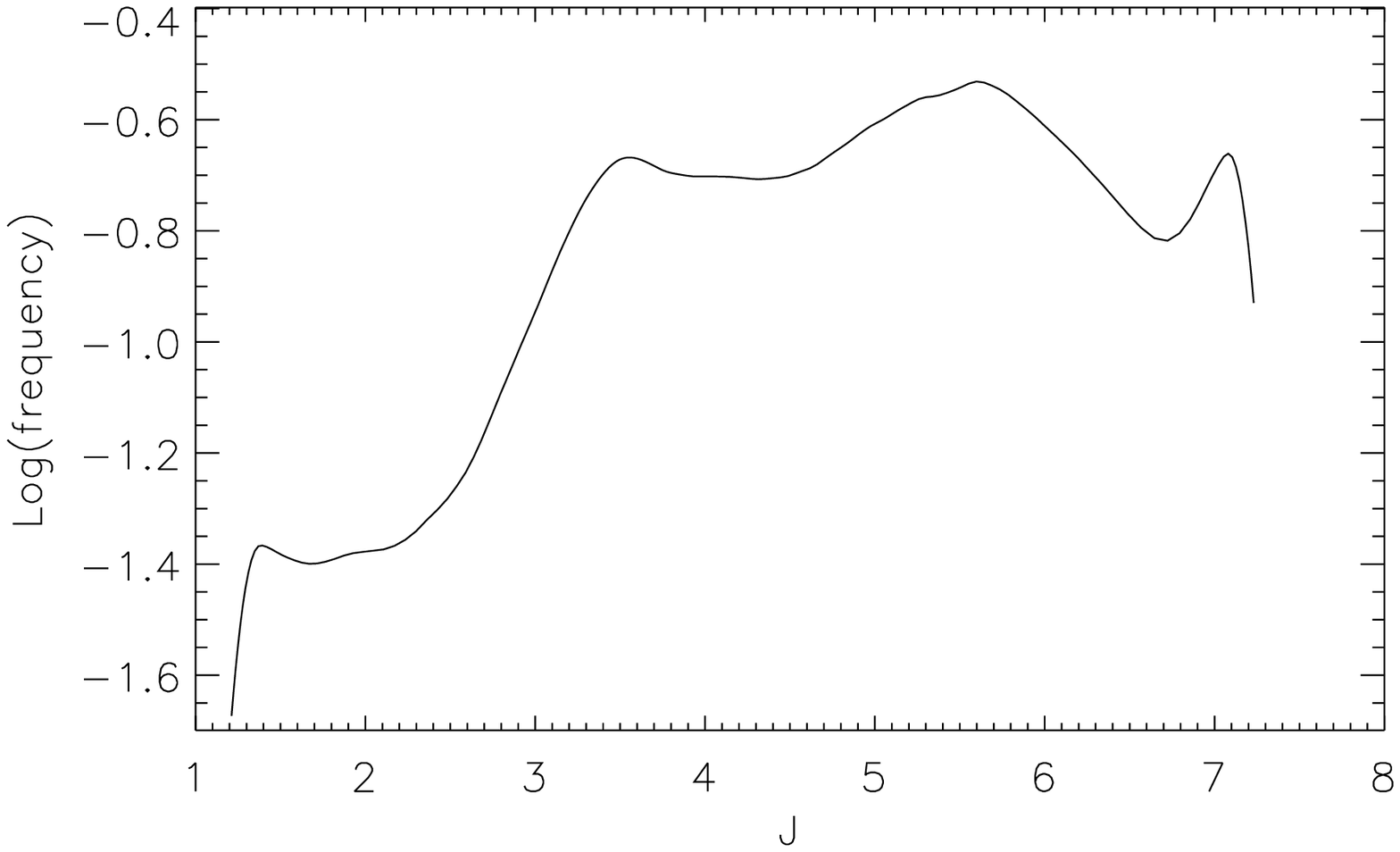}
\caption{\textit{Top panel - }Mass--\teff\ relation provided by the models of DM98, DUSTY and SDF00. We also show our smooth relation. \textit{Bottom panel - }Luminosity function derived from the mass distribution of \citet{Muench02}, and assuming our semi-empirical mass--\teff\ relation shown in the top panel.}\label{fig:mass_teff}
\end{figure}

Our empirical isochrone performs the second step of the process, converting \teff\ into magnitudes, but does not provide any information on the first passage, the mass vs \teff\ relation. To this purpose, we derive an average mass--\teff\ relation by combining the relations of the different theoretical models (Fig.~\ref{fig:mass_teff}).

\begin{figure}
\centering
\includegraphics[width=\linewidth]{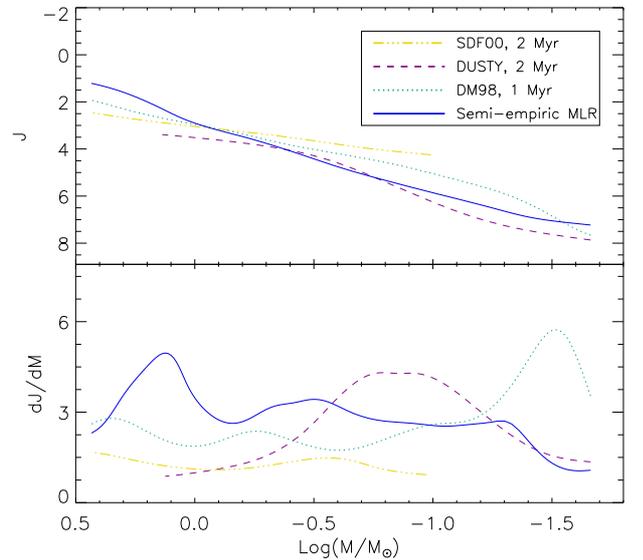}
\caption{Smoothed MLRs (top panel) and the corresponding derivatives (bottom panel) for the theoretical models discussed in the text.}\label{fig:mlrs}
\end{figure}

\begin{figure}
\centering
\includegraphics[width=\linewidth]{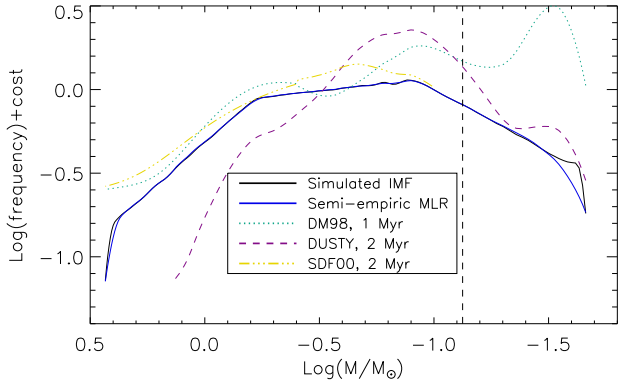}
\caption{Comparison of the IMFs derived using different MLRs. The solid black line is the mass distribution of our artificial sample. The dashed vertical line marks the separation between stars and brown dwarfs ($M=0.075M_\odot$).}\label{fig:imf}
\end{figure}

Combining the mass--\teff\ relation with the empirical isochrone listed in Table~\ref{tab:colors} we obtain our semi-empirical MLR (Fig.~\ref{fig:mlrs}) needed in Eq.~\ref{eq:imf}.  We extend it to spectral types earlier than K6 using the 1~Myr old model of DM98, which smoothly connects to our isochrone. The \jb-LF of our artificial sample of stars is shown in Fig.~\ref{fig:mass_teff}.

The last passage of our exercise requires converting the LF in Fig.~\ref{fig:mass_teff} back into a mass function by means of Eq.~\ref{eq:imf}, assuming different MLRs\footnote{We remind the reader that the luminosity spread discussed in Sect.~\ref{sec:empiric} prevents any luminosity-mass conversion on a star-by-star basis. Nonetheless, this statistical approach remains valid when using the most appropriate MLR on a large sample of stars, as the errors should average down.}.
As a sanity check, we start with our semi-empirical MLR (Fig.~\ref{fig:mlrs}), used to generate the LF. Figure~\ref{fig:imf} shows that, besides some numerical oscillation, we recover our assumed mass distribution. This proves the robustness of our algorithm. We then perform the same computation assuming the \jb-band MLRs provided by the models discussed in Sect.~\ref{sec:discussion}. We apply a kernel smoother to the MLRs in order to reduce the numerical fluctuations induced by  the limited and uneven mass samplings of the model relations (Fig.~\ref{fig:mlrs}), obtaining the results presented in Fig.~\ref{fig:imf}.

Let's focus first on the stellar regime. Using the theoretical models, we generally recover the main features of the simulated IMF: it increases with decreasing masses for $M>0.6M_\odot$ and then flattens and reaches the maximum frequency. The recovered peak location depends on the assumed model: the DM98 and DUSTY models return the same characteristic mass ($M=0.12_\odot$) as the simulated IMF, while the model of SDF00 returns a slightly larger mass ($M\sim0.23 M_\odot$) corresponding to the peak position.

Following \citet{Muench02}, we fit the IMFs using the power-law
\begin{displaymath}
\frac{dN}{dM_L}=M^\Gamma
\end{displaymath}
with turn-off mass at $M=0.6 M_\odot$. In Table~\ref{tab:gamma} we list the derived $\Gamma$ coefficients: we find that using the theoretical MLRs we overestimate the simulated $\Gamma$ coefficient in both mass ranges. In particular, the DUSTY model shows the largest deviation from the expected results. This is due to the fact that while the derivatives of the MLR provided by the other models are roughly constant in the $M<0.12 M_\odot$, the derivative corresponding to the DUSTY model increases with decreasing mass.

\begin{table}
  \centering
  \caption{Fit coefficients $\Gamma$ derived assuming different model MLR }\label{tab:gamma}
  \begin{tabular}{c|cccc}
    \hline\hline
 & Simulated & DM98 & DUSTY & SDF00\\
    \hline
$M>0.6 M_\odot$ & -0.78\tablefootmark{a} & -1.07 & -2.19 & -0.77\\
$0.12M_\odot<M<0.6 M_\odot$ & -0.15\tablefootmark{a} & -0.26 & -1.11 & -0.25\\
    \hline
    \end{tabular}
\tablefoot{
   \tablefoottext{a}{From \citet{Muench02}.}
}
\end{table}

Considering the substellar domain, we find that both the DUSTY and the DM98 models provide an over-population of brown dwarfs, both of them generating a feature in the IMF (as the one seen in the ONC IMF provided by \citet{Muench02} or \citet{LR00}) which could be mistaken as a secondary peak in the substellar regime.

Splitting the IMFs in two parts around the peak ($M=0.12 M_\odot$) and integrating in the corresponding mass ranges, we find that the fraction of very low-mass objects ($M<0.12 M_\odot$) compared to the number of stars with $0.12M_\odot<M<1M_\odot$ is 1.4 and 0.7, assuming the DM98 and DUSTY models respectively. These fractions are inconsistent with the true population of stars, which provides a 0.4 fraction. This because the two evolutionary models provide luminosities which decrease too rapidly with the stellar mass (see Fig.~\ref{fig:hr}) and, by consequence, the derivatives of the MLRs in the substellar regime are higher than our semi-empirical expectation. 

For the sake of comparison with observational results, we remark that our Fig.~\ref{fig:imf} is similar to Fig.~18 in \citet{Muench02}, where the authors summarize a few derivations of the IMF for the central 5\arcmin surrounding the Trapezium stars. Such a small area is projected against the thickest part of the background Orion Molecular Cloud \citep[$A_V\gtrsim25$, ][]{Scandariato10}, thus galactic background contamination is negligible. As a consequence, differences between the observationally-derived IMFs are due to differences between techniques (photometry, spectroscopy\dots) and assumed theoretical models. From this comparison it is evident that at the low-mass end, the IMF derivations appear to diverge, and it is unclear how to make detailed comparisons between the different methods. In any case, since photometrically-derived IMFs primarily depend on the theoretical MLR extracted from the PMS model, the differences between them likely stem from the inaccuracies in the evolutionary models at subsolar masses, as we show in Fig~\ref{fig:imf}.

\section{Conclusion}\label{sec:conclusion}

In this paper we have accurately analyzed a sample of $\sim$300 stars with measured temperatures, luminosities and photospheric NIR photometry as a benchmark for current atmospheric and evolutionary models for low-mass PMS stars and brown dwarfs.

We have compared the extinction-corrected photometry to the expected photometry provided by the template spectra of \citet{Allard10}, finding that major improvements have been done in the synthesis of theoretical spectra. Nonetheless, we obtain indications on the lack of opacity in the \hb-band, likely due to the improper treatment of either the water vapor absorption profile or the collision induced absorption from H$_2$. We thus propose the set of empirical corrections listed in Table~\ref{tab:atm}, to be regarded as additive terms to the synthetic NIR colors. These corrections are weakly influenced by the $\log g$ assumed to derive synthetic colors.

We also analyzed the same sample of stars in order to derive the average isochrone of the ONC, reported in Table~\ref{tab:colors}. The analyzed sample of stars show a magnitude spread of the order of $\sim$3.5~mag, consistent the the $\sim$1.5~dex luminosity spread reported in previous study of the ONC. This spread does not allow us to constrain the magnitude scale of the ONC isochrone and, by consequence, does not allow any mass determination based solely on the magnitude of stars. On the other hand, the \jb-\hb\ and \hb-\kb\ colors weakly depend on the photospheric luminosity, and are well constrained by our statistical analysis.

Comparing our empirical isochrone to current theoretical models, we find that there is generally good agreement with the 2~Myr old model of DUSTY and, to some extent, the 2~Myr old model of SDF00, both in magnitude (down to M5 types) and color scale. On the other hand, we find indications that the PMS isochrones of DM98 provide \teff--$\log L$ relations slightly flatter than our observational data.

Finally, we investigated how the theoretical models affect the photometric derivation of the IMF in the NIR domain. We find that PMS evolutionary models generally underestimate the intrinsic luminosity of VLMSs and BDs, and this may lead to artificial structures in the low-mass tail of young clusters' mass distribution.

The empirical NIR colors we have derived, in the 2MASS system, can be readily converted to other  photometric systems using the prescriptions at \url{http://www.ipac.caltech.edu/2mass/releases/allsky/doc/sec6_4b.html}. Since low-mass stars slowly evolve in the HR diagram at very early stages, our empirical colors can conveniently be used as the intrinsic colors of young ($age\lesssim5$~Myr) stellar clusters to derive NIR excess and extinction of individual stars.

\begin{acknowledgement}
This research has benefitted from the SpeX Prism Spectral Libraries, maintained by Adam Burgasser at \url{http://pono.ucsd.edu/~adam/browndwarfs/spexprism}.
\end{acknowledgement}

\end{document}